\documentstyle[12pt,epsf]{article}
\begin{document}
\title{
{\bf The energy behaviour of real and virtual photon--proton cross
sections}
 \footnote{Based on a talk given at the fifth Gentner Symposium
on Physics, Dresden, 16--21 October, 1994}\\
 \author{
{\bf Aharon Levy}\\
DESY, Hamburg, Germany
\\
and \\
Tel--Aviv University, Tel--Aviv, Israel
}
 }
\date{ }
\maketitle
\vspace{5 cm}
\begin{abstract}
The recent F$_2$ measurements at HERA are discussed in the framework of
virtual photon--proton cross section. The energy behaviour of the cross
section is studied for different $Q^2$ values, ranging from 0 to 1000
GeV$^2$ for center of mass energies $1.75 < W < 300$ GeV.
\end{abstract}
\vspace{-20cm}
\begin{flushleft}
\tt DESY 95-003 \\
January 1995\\
\end{flushleft}

\newpage

\section{Introduction}

The recent measurements of $F_2$ at HERA by the H1~\cite{h1}
and ZEUS~\cite{zeus}
collaborations open a new kinematic domain for the study of the
structure function of the proton. The new HERA data showed a dramatic increase
of $F_2$ with decreasing $x$. These measurements can be interpreted as virtual
photon--proton cross sections. The $F_2$ data (for $Q^2 \neq$ 0) can be
converted into $\sigma_{tot}(\gamma^*p)$ as a function of $W$ using the
relation:
\begin{equation}
\sigma_{tot}(\gamma^*p) =
\frac{4\pi^2\alpha}{Q^4}\frac{4m_p^2x^2+Q^2}{1-x}F_2(x,Q^2)
\end{equation}
and using
\begin{equation}
W^2 = m_p^2 + Q^2 (\frac{1}{x} -1)
\end{equation}

The ZEUS collaboration showed that when their $F_2$ data are interpreted as
total virtual photon--proton cross sections, $\sigma_{tot}(\gamma^*p)$ has a
fast rise with $W$ in the range $50 < W < 280$ GeV for $ 8.5 < Q^2 < 125$
GeV$^2$. In this note a compilation of cross
sections~\cite{zeus,bcdms,slac,nmc,na28,e665} is presented for a wider range of
$Q^2$ and $W^2$ to study the behaviour of the total cross section as function
of
both variables.

\section{Behaviour of $\sigma_{tot}(\gamma^*p)$ for $W^2 < 400$ GeV$^2$}

Earlier studies of the behaviour of $\sigma_{tot}(\gamma^*p)$ with $W$
were limited to center of mass energies $W <$ 20 GeV. In
{}~\cite{lm} we studied the data of the EMC collaboration~\cite{emc} DIS
$\mu$--p. It was shown that when the $\gamma^*p$ cross sections are
plotted as a function of $W^2$, for fixed $Q^2$, they are strongly
increasing with $W$ for high $Q^2$, while they have a much milder
dependence on $W$ for low $Q^2$. The strong rise at high $Q^2$ can be
associated with the threshold behaviour of a cross section corresponding
to a virtual particle with $Q^2 > W^2$. Figure 1 is taken from ~\cite{lm}
and shows the EMC data together with curves which are a fit to a Regge
type energy behaviour to the data. The fitted expression also includes a
threshold factor to explain the sharp rise with $W$ at higher $Q^2$ values.

\section{Behaviour of $\sigma_{tot}(\gamma^*p)$ for $W^2 < 10^6$ GeV$^2$}

Data are shown in figure 2 for $Q^2$ values starting from $Q^2$=0 (real
photoproduction) up
to $Q^2=1000$ GeV$^2$. The energy range is $3 < W^2 < 90000$ GeV$^2$,
where the lower limit is chosen so as to be above the resonance
region. In order not to overcrowd the picture, selected $Q^2$ regions are
shown.

The following observations can be pointed out from the behaviour of the data.
In
the low $W$ region ($W^2 <$ 400 GeV$^2$) one observes a slow decrease of the
cross
section with energy for the lower $Q^2$ region. As $Q^2$ increases, one starts
to observe the threshold effects shown in figure 1 for cases where $Q^2 > W^2$.
These cause the cross section to rise steeper as $Q^2$ increases.

In the higher $W$ range, there exist now the new total photoproduction
cross section measurements for $Q^2 \approx$ 0 by ZEUS ~\cite{zeustot}
and by H1 ~\cite{h1tot} which show that the cross section has a slow
increase with energy, in accord with expectations from Regge theory based
models ~\cite{dl,allm}. There are no measurements in the low $Q^2$ range
($\sim$ 0.3-5 GeV$^2$) at high $W$, so one cannot see if it behaves in the
same way as the $Q^2$ = 0 photoproduction data. However as $Q^2$
increase ($> 8-10$ GeV$^2$) there is a definite change of the energy
behaviour of $\sigma_{tot}(\gamma^*p)$. The energy dependence becomes much
steeper than at low $Q^2$.

The curves in figure 2 are the calculated $\gamma^*p$ cross sections using
the ALLM~\cite{allm} parametrization. The parameters of ALLM have been
obtained by a fit to the available data below $W =$ 20 GeV. It is a
remarkable fact that the low $W$ data constrained the parameters so as to
force a rising cross section at high $W$ and high $Q^2$. This is
connected to the fact that the Pomeron trajectory was allowed to vary
with $Q^2$, as shown in figure 3. The exact location of the transition
from values of $\alpha_P(0) \simeq$ 1.05 to $\alpha_P(0) \simeq$ 1.4 is
not clear since there is a lack of data in the region of $1 < Q^2 < 10$
GeV$^2$ and $W >$ 20 GeV.
The ZEUS data indicate that the transition from a slow to a steeply rising
Pomeron is more gradual than proposed by figure 3 and even at a $Q^2$ of $\sim$
15 GeV$^2$ the slope is still halfway between the two extreme values.
The shape assumed by the ALLM parametrization
in the transition region could account for the fact that the curves don't
reproduce the high $W$ data for the $Q^2$ values of 8.7 and 15 GeV$^2$,
shown in figure 2. It does however fit quite well both the $Q^2$ = 0 and
the higher $Q^2$ data in the whole $W$ region.

In order to get a better understanding of the data in figure 2, one clearly
needs more measurements to fill in two large gaps ~\cite{frank}. One gap is the
low to intermediate $Q^2$ region, where the transition of the energy dependence
from a slow to steep rise occurs. In these region higher $W$ measurements are
needed, which would correspond to lower $x$ measurements at low $Q^2$.
The data of E--665~\cite{e665}, once in final form, will fill up part of this
gap.
The other
gap is the energy range of $W^2$ = 400--40000 GeV$^2$ at higher $Q^2$ values.
This would correspond to DIS measurements at higher $x$. These additional
needed
measurements would cover the holes in the $x$--$Q^2$ plane and will provide
additional handles to study the transition region between the so--called soft
Pomeron with a Donnachie--Landshoff type of slope of 1.08 and the hard Pomeron
with a Lipatov type of slope of about 1.5.

\section{One schizophrenic Pomeron or two different Pomerons?}

The interpretation of the $Q^2$ dependence of the Pomeron intercept shown
in figure 3 could be that the Pomeron trajectory has a different
behaviour than that of the other known ones, changing its intercept as
$Q^2$ increases. This interpretation is of course not unique. As pointed
out by Bjorken ~\cite{bj}, another possibility could be that the
intercept plotted in figure 3 is an effective one, being the
superposition of two different Pomeron trajectories, one having an
intercept of 1.08 and the other one of 1.5, as illustrated in figure 4. The
slope of the $t$ dependence is usually taken as 0.25 GeV$^{-2}$ for the soft
Pomeron. For the Hard Pomeron, the slope shown is arbitrarily taken as 0.025
GeV$^{-2}$. The amount of mixture of the two intercepts is $Q^2$ dependent:
\begin{equation}
\alpha_P^{eff} = a(Q^2) \alpha_{P_{soft}} + (1 - a(Q^2)) \alpha_{P_{hard}}
\end{equation}
with the dominant contribution of the
first in the low $Q^2$ region and the dominance of the second in the high
$Q^2$ region.

Continuing in the same speculative spirit, one could ask oneself about the
strong resemblance between the behaviour of the Pomeron(s) and the photon. Do
we see a 'resolved' and 'direct' Pomeron like in the photon case? Any relation
between these 'components' and the soft and hard Pomerons?

The hope is that the expected new measurements of the two
HERA collaborations in the 'gap' regions will allow a better
understanding of the transition region and may shed some light on this so
important issue of the Pomeron properties.

\section{Behaviour of $\sigma_{tot}(\gamma^*p)$ for $W^2 > 10^6$ GeV$^2$}

Another question that one would like to study is whether the total $\gamma^*p$
will continue to rise for $W^2 > 10^6$ GeV$^2$ ~\cite{bjsum}. Will there be a
cross--over of
virtual and real photon--proton cross sections or will $\sigma_{tot}(\gamma p)$
also start rising more steeply, as predicted by some QCD inspired models? One
way to reach this high $W^2$ region would be to wait for the $e$--p option of
the LHC. However as an alternative to waiting at least 20 years for an answer,
one should perhaps consider the possibilities of raising the
energies at the HERA collider. Since
\begin{equation}
W^2 \simeq y s \simeq 4 y E_e E_p,
\end{equation}
increasing the energy of the electron beam, of the proton beam or perhaps of
both, will provide the answer as far as the rise is concerned. In addition, it
will get us to much lower $x$ regions ($\sim 10^{-6}$) than at present
energies, allowing to study parton saturation effects.

\section{Summary}

As strange as it might sound, the understanding of the complicated nature of
the
Pomeron is necessary to improve our picture of QCD. One can study the
interesting properties of the Pomeron through the measurements of the total
cross sections of real and virtual photons, which show a change in their energy
behaviour by going from real to highly virtual photons. In order to learn more
about the transition region one needs to fill some gaps in the measurements of
the total $\gamma^*p$ cross section at low $Q^2$ and intermediate $W$. This
will be achieved in the near future. In addition, one needs to measure both
$\sigma(\gamma p)$ and $\sigma(\gamma^*p)$ at  $W$'s larger than presently
available at HERA.


\section{Acknowledgements}

It is a pleasure to acknowledge useful discussions with H.Abramowicz.
Thanks are due also to R.Klanner and J.Whitmore for a careful reading of the
text.
This work was partly supported by the German Israeli Foundation (GIF) and by
the
Israel Academy of Science.

\newpage
\begin{figure}[h]
\includegraphics{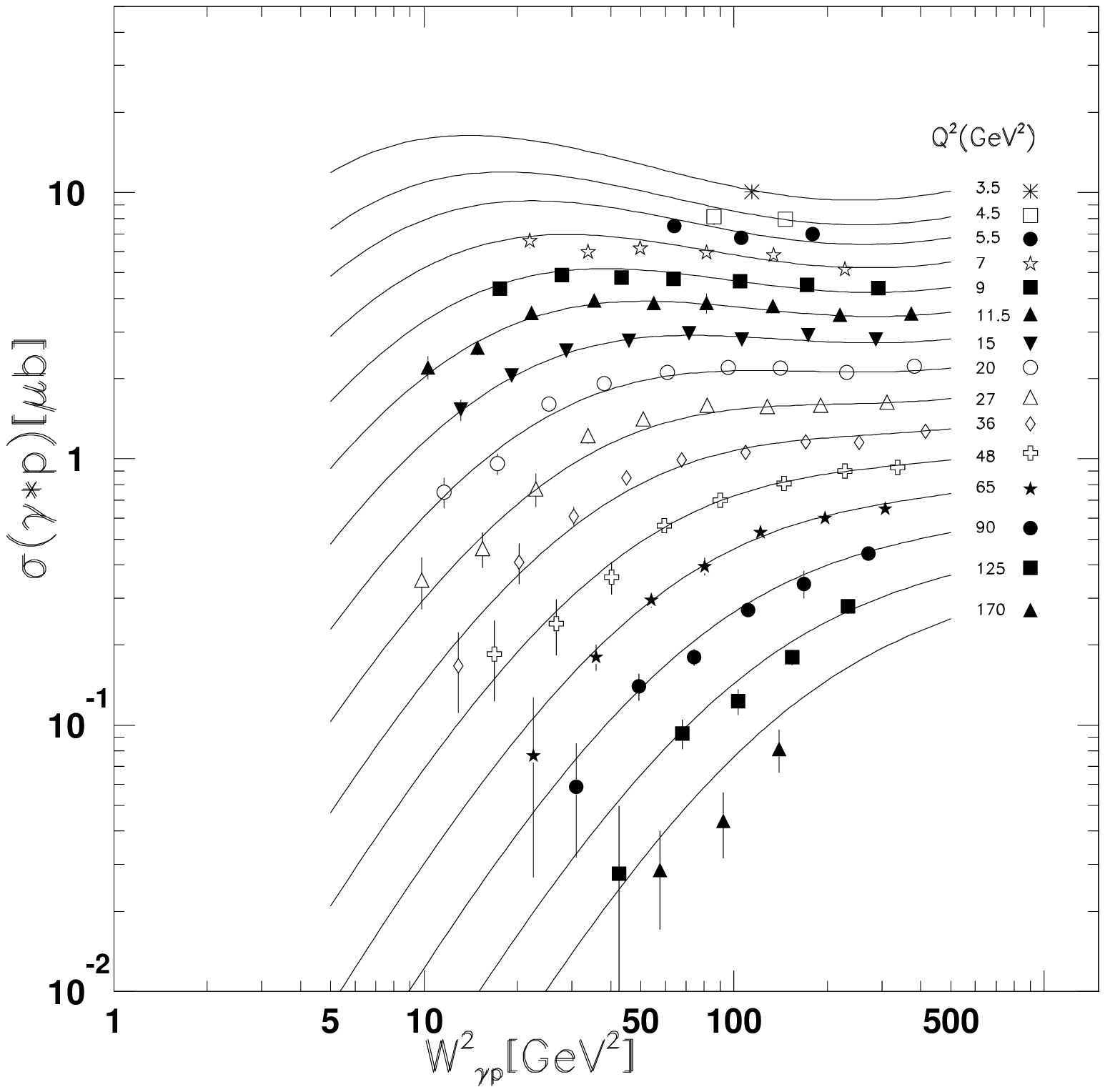}
\unitlength1cm
\begin{picture}(15,20)
\thicklines
\end{picture}
\caption{ Sample of $\gamma^*p$ total cross section data points
 from [4] and the fits of [3].}
 \label{fig1:lmfig}
\end{figure}

\newpage
\begin{figure}[h]
\includegraphics{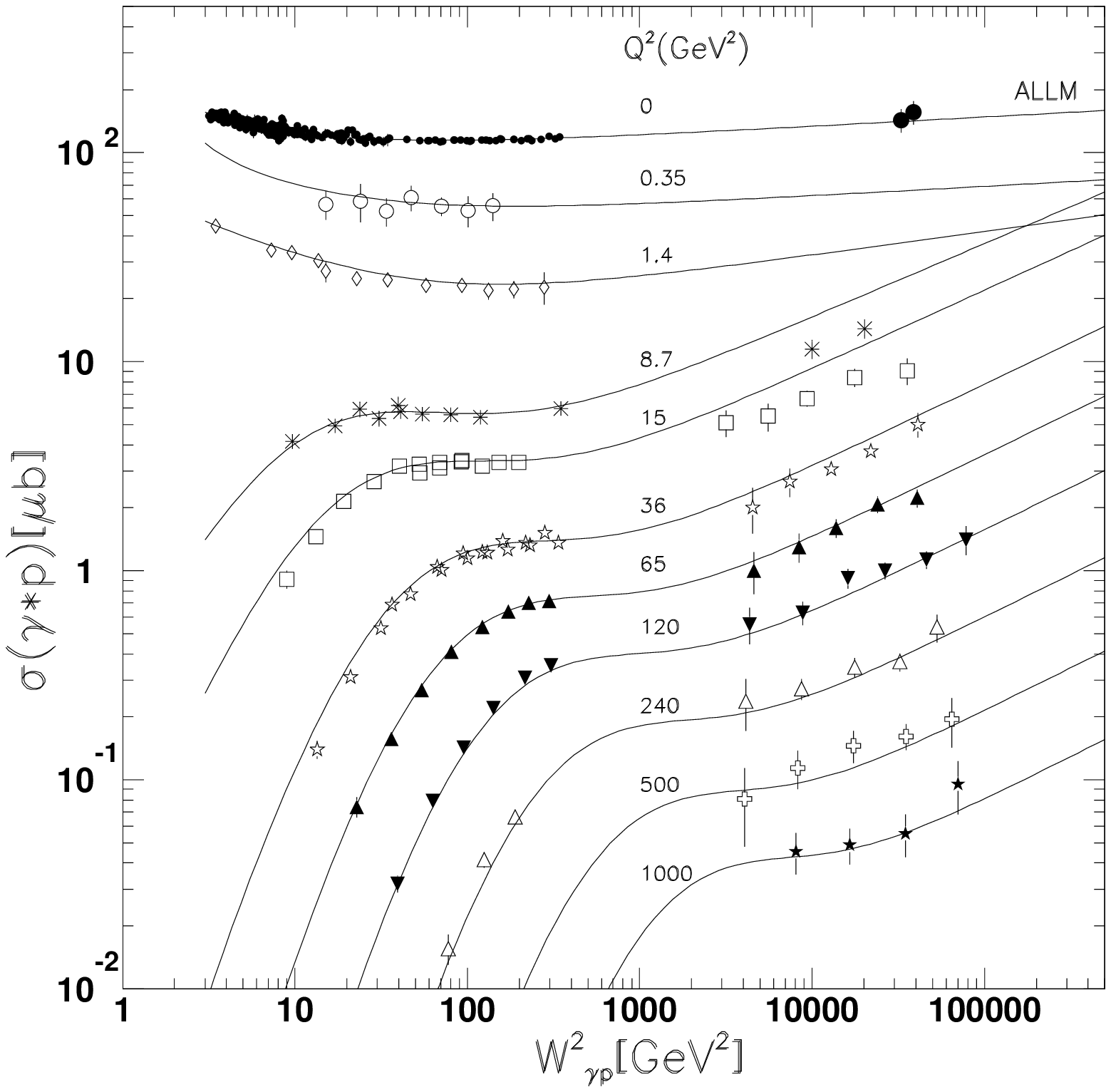}
\unitlength1cm
\begin{picture}(15,20)
\thicklines
\end{picture}
\caption{ Total $\gamma*p$ cross section as function of the center of
mass energy squared, $W^2$. The curves are the ALLM parametrisation,
fitted to the lower energy data ($W^2 <$ 400 GeV$^2$), and extrapolated to the
high energy region.} \end{figure}

\newpage
\begin{figure}[h]
\includegraphics{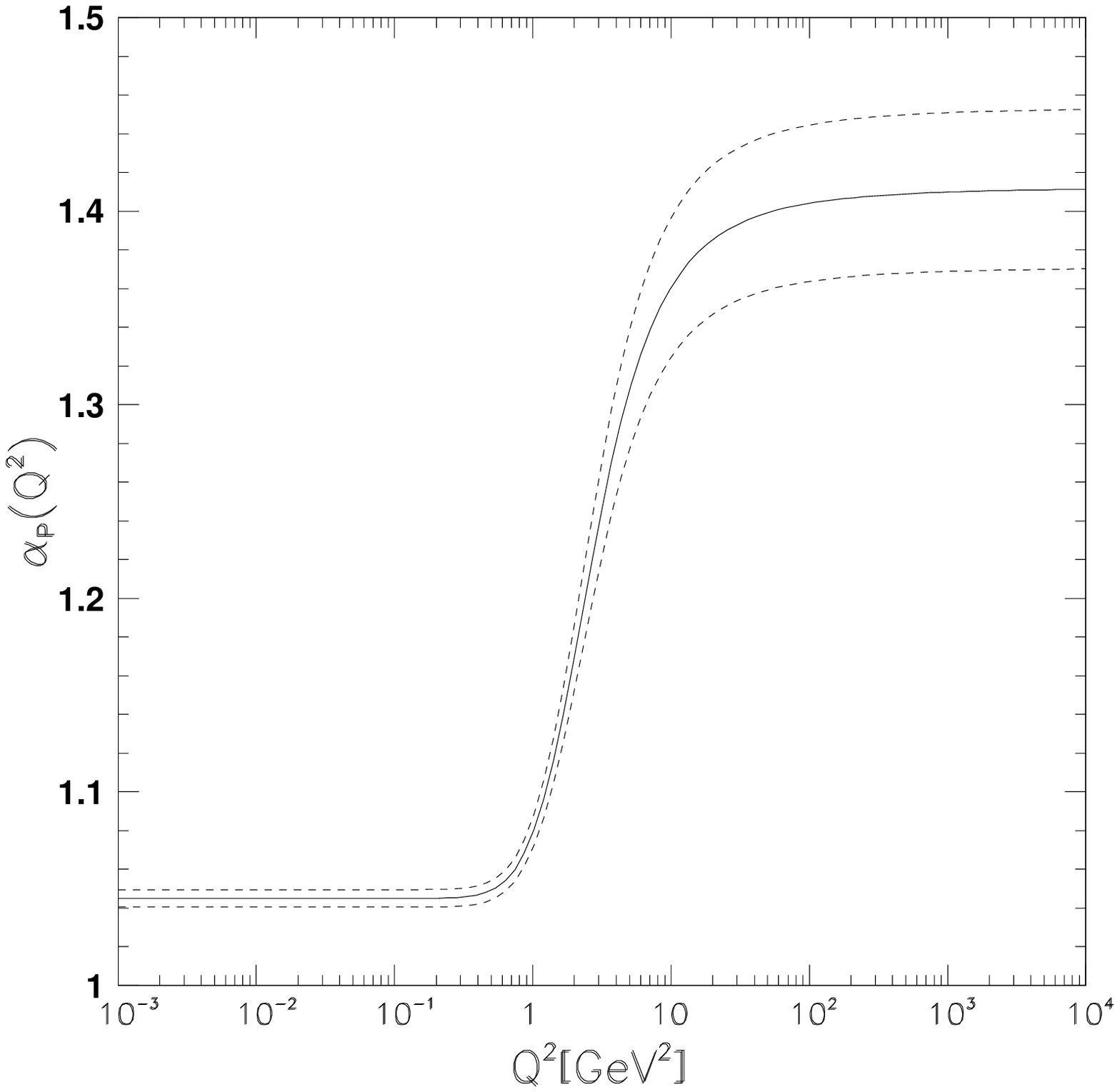}
\unitlength1cm
\begin{picture}(15,20)
\thicklines
\end{picture}
\caption{The intercept of the Pomeron trajectory $\alpha_P(0)$ as
function of $Q^2$, as obtained from the ALLM parametrisation. The dotted
line show the uncertainty of the fit.}
\end{figure}

\newpage
\begin{figure}[h]
\includegraphics{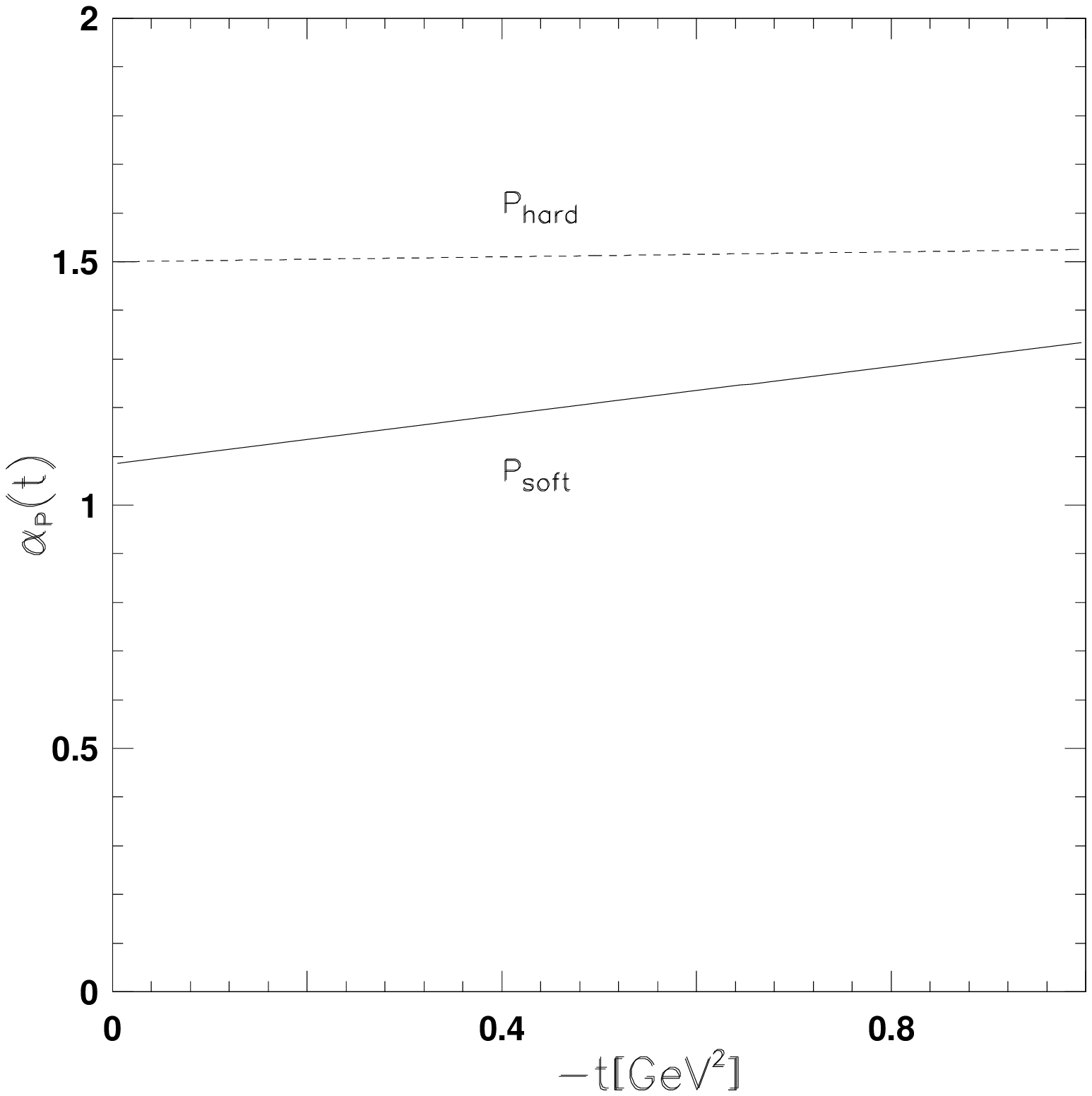}
\unitlength1cm
\begin{picture}(15,20)
\thicklines
\end{picture}
\caption{The trajectory of the
soft Pomeron, $P_{soft}$ and that of the hard Pomeron, $P_{hard}$, as function
of t.} \end{figure}


\newpage

\begin{thebibliography}{50}

\bibitem{h1} H1 Collab., Preliminary results presented by J.Feltesse,
Rapporteur talk at the 24th International Conference on High Energy Physics,
Glasgow, July 1994.
\bibitem{zeus} ZEUS Collab., M.Derrick et al., DESY 94--143.
\bibitem{lm} A.Levy, U.Maor, Phys. Lett. {\bf B182} (1986)
108. \bibitem{emc} EMC Collab., J.JAubert et al., Nucl. Phys. {\bf B259}
(1985) 189.
\bibitem{bcdms} BCDMS Collab., A.C.Benvenuti et al., Phys. Lett. {\bf B223}
(1989) 485.
\bibitem{slac} SLAC Collab., L.W.Whitlow et al., Phys. Lett. {\bf B282} (1992)
475.
\bibitem{nmc} NMC Collab., D.Allasia et al., Phys. Lett. {\bf B249} (1990) 366.
\bibitem{na28} NA28 Collab., M.Arneodo, et al., Nucl. Phys. {\bf B333} (1990)
1.
\bibitem{e665} E665 Collab., Preliminary results presented by H.Schellman at a
DESY seminar, Sep 1994.
 \bibitem{zeustot} ZEUS Collab., M.Derrick et al., Z. Phys. {\bf C63}
(1994) 391; Phys. Lett. {\bf B293} (1992) 465.
\bibitem{h1tot} H1 Collab., T.Ahmed et al., Phys. Lett. {\bf B299} (1993) 374.
\bibitem{dl} A.Donnachie, P.V.Landshoff, Nucl. Phys. {\bf B244} (1984) 322.
\bibitem{allm} H.Abramowicz et al., Phys. Lett. {\bf B269} (1991) 465.
\bibitem{frank} F. Sciulli, Summary talk of the International Workshop on DIS,
Eilat, Israel, Feb 1994.
\bibitem{bj} J.D.Bjorken, Proceedings of the International
Workshop on DIS, Eilat, Israel, Feb 1994.
\bibitem{bjsum} J.D.Bjorken, Summary talk of the International Symposium on
Multiparticle Dynamics, Salerno, Italy, Sep 1994.
\end{thebibliography}
 \end{document}